\documentclass[aps, twocolumn]{revtex4-1}

\usepackage{graphicx}
\usepackage{amsmath}
\usepackage{dsfont}
\usepackage{amssymb}
\usepackage{physics}
\usepackage{hyperref}
\usepackage{ulem}
\hypersetup{colorlinks=true,
	    final=true,
	    linkcolor=blue,
	    citecolor=blue,
	    filecolor=blue,
	    urlcolor=blue,}

\newcommand{\pr}{\partial}

\newcommand{\ep}{\epsilon}

\newcommand{\p}{\prime}
\newcommand{\om}{\omega}
\newcommand{\ra}{\rangle}
\newcommand{\la}{\langle}

\newcommand{\beq}{\begin{equation}}
\newcommand{\eeq}{\end{equation}}

\newcommand{\mathq}{\mathbf{q}}
\newcommand{\mathk}{\mathbf{k}}
\newcommand{\mathkp}{\mathbf{k^\p}}

\begin{document}

\title{Hot (non-equilibrium) electron relaxation: A review of the ultra-fast phenomena in metals and superconductors (PART I)}
\author{Navinder Singh}
\email{navinder.phy@gmail.com; Phone: +91 9662680605}
\affiliation{Theoretical Physics Division, Physical Research Laboratory (PRL), Ahmedabad, India. PIN: 380009.}

\begin{abstract}
The famous Two-Temperature Model (TTM) used extensively in the investigations of energy relaxation in photo-excited systems originated in the seminal work of M. I. Kaganov, I. M. Lifshitz, and L. V. Tanatarov (KLT) in 1957. The idea that with an ultra short laser pulse a temporal (transient) state of electrons in a metal can be created in which electrons after absorbing energy from the laser pulse heat up and their temperature becomes substantially greater than that of lattice was originated in the work of S. I. Anisimov, B. L. Kapeliovich, and T. L. Perel'man in 1974.  The heated electron sub-system ("hot" electrons) loses its energy to phonon sub-system via electron-phonon scattering (relaxation) and thermodynamic equilibrium re-establishes over a time scale of a few pico-seconds (psecs) in metals. This field saw great developments in the 1980s and 1990s with the advent of femto-second (fsec) pump-probe spectroscopy. And from 2000 onwards focus shifted from non-equilibrium phenomena in simple metals to that in more complex systems including strongly correlated systems such as high Tc cuprate superconductors. In 1987, P. B. Allen re-visits the calculations of KLT and re-writes the electron-phonon heat transfer coefficient $\alpha$ in terms of a very important parameter in the theory of superconductivity ($\lambda \la \om^2\ra$). This has far-reaching consequences, $\lambda$, a very crucial parameter for a given superconducting material, can be estimated by doing pump-probe experiments on it. By mid 1990s, it became clear that TTM is violated and is not a sufficient model to discuss non-equilibrium relaxation. Year 2000 onwards, field saw the development of models that go beyond the original TTM. Very recently, field enters into atto-second domain. In this article, author attempts a rigorous and concise account of the entire development of this very fascinating area of research including a summary of the current research in this field.
\end{abstract}

\maketitle

\texttt{ -- One of the disadvantages of ``scientific fashions"  is that they sometimes displace old but important scientific problems. --}

 \section{Author's note on the presentation} 

 This series of two articles covers a considerable ground of this very interesting and important field of research (starting from the pioneering work of M. I. Kaganov, I. M. Lifshitz, and L. V. Tanatarov in 1957 to the application of P. B. Allen's ideas to investigate the "pairing glue" in high Tc cuprates in 2012 and other recent investigations). The author, however, skipped the topic of non-equilibrium electron relaxation in nano-systems (nano-particles and thin films). This topic is covered to some extent in author's older review article\cite{nav1}. This first article (part I) discusses the ultrafast phenomena in metals, and the second article (Part II) will discuss that phenomena in superconductors (both conventional and unconventional). The presentation is designed to give the reader a comprehensive overview (a perspective) of the field, ensuring they don't get lost in the details (i.e., helping them avoid ``missing the wood for the trees").   Regarding way of presentation, a common question that arises in an author's mind when preparing a review of a scientific field is whether to adopt a chronological presentation of the development of ideas or to present them in a logical manner that may not necessarily follow a chronological path. The author reflected on this while planning to write a review of this fascinating field of research. He believes that a chronological presentation is, in fact, the best way for this field, and thus has chosen to follow this approach. Referencing is no way complete, although author has tried his best to cite important works which significantly contributed to the understanding of the field. The author apologizes at the outset for missing some important works. Any important references that were skipped may be brought to his attention.

 \begin{figure}[h!]
    \centering
    \includegraphics[width=1.0\columnwidth]{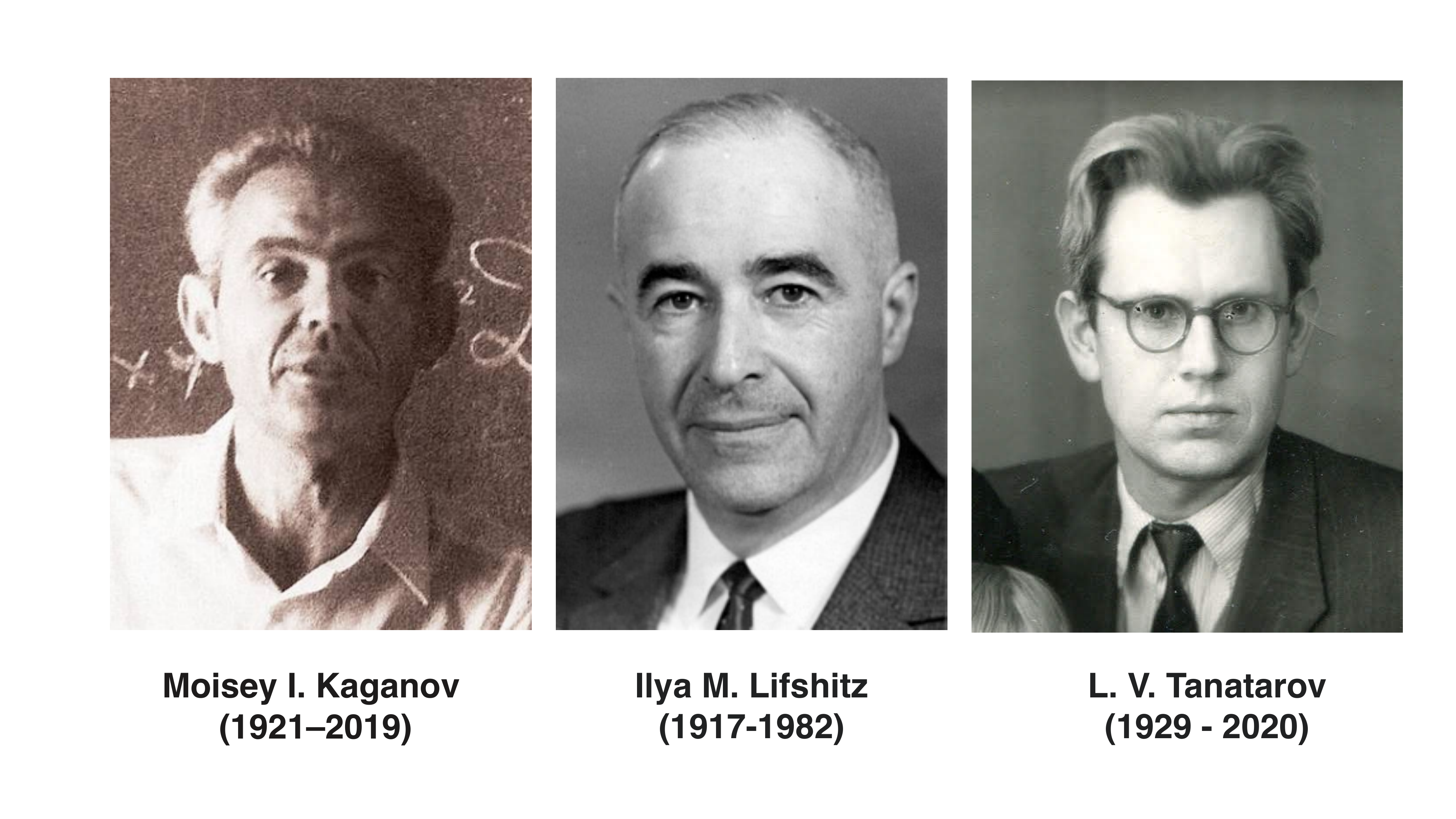}
    \caption{Through this article author pays his tribute to the pioneers of the Two-Temperature model. Image courtesy: M. I Kaganov (from Yaroslaw Bazaliy via peronal communication);  I. M. Lifshitz (Wikipedia Commons); L. V. Tanatarov (from Igor Tanatarov (grandson of L. V. Tanatarov)  via personal communication).}
    \label{f1}
\end{figure} 

\section{A bit of history}

The idea of the two-temperature model of hot electron relaxation in metals originated in the studies of radiation damage in metals by very high energy ions, when metals are exposed to such ions. It all started near the end of World War II in 1945 and such studies were carried out in the so called ``the Laboratory No. 1" in the Ukrainian Physico--Technical Institute (UPTI) in Kharkiv\cite{baz}. This section was then headed by O.I. Akhiezer and I. M. Lifshitz and Moisei Isaakovych Kaganov joined in to work on these projects related to radiation damage in metals by high energy ions and other topics in metal physics\cite{histo}. For several years, the results were not announced (these were kept classified) and finally some of the results were announced in 1957\cite{klt}. The understanding that originated from those studies can be expressed in the following way. It was pointed out that the damage in a metal by high energy ions occurs through a series of processes in cascade and various relaxation process are separated from each other in time. High energy ions as they penetrate in a metal, first transfer their energy to electrons as electrons has much less specific heat as compared to lattice. Due to this, electron sub-system heats up (preferentially) and internal electron-electron scattering leads to a "hot" Fermi-Dirac distribution of electrons at an elevated temperature. This "hot" electron distribution then transfers its energy to phonon sub-system via slower electron-phonon scattering. They argued that electron-electron relaxation is much faster as compared to electron-lattice relaxation. And during the electron-phonon scattering electron distribution remains in equilibrium Fermi-Dirac distribution but an elevated temperature.

The first method used by them is classical one in which radiation of sound waves by a fast moving electron through the lattice ("Cerenkov" radiation) is computed by considering excitation of vibrations of an elastic continuum (on the lines of a method developed by L. D. Landau\cite{lan}) . The second method, which is fully quantum, is an extension of a beautiful set of calculations by A. I. Akhiezer and I. Pomeranchuk\cite{akh}. Akhiezer and Pomeranchuk consider the mechanism of spin-lattice relaxation in the context of magnetic method of cooling and used the Bloch-Boltzmann kinetic equation. M. I. Kaganov, I. M. Lifshitz, and L. V. Tanatarov (figure 1)  applied the Akhiezer-Pomeranchuk method for the computation of relaxation time between "hot" electrons and lattice. The idea of laser excitation of electrons was not there at that time. In fact laser was not discovered at that time (it came only in 1960 due to investigations of Theodore Maiman and others).  These authors considered disequilibrium between electrons and lattice as arising due to the passage of high energy ions in metals, and also when large current is passed through a metal, such that Ohm's is violated\cite{klt}. In the following section we review the Two-Temperature Model (TTM).

\section{The Two-Temperature model (TTM)}

Kaganov-Lifshitz-Tanatarov (KLT)\cite{klt} assumed that after preferential heating, electrons quickly regain the Fermi-Dirac distribution (via electron-electron collisions) albeit at an elevated temperature (that is "hot" Fermi-Dirac distribution):
\beq
f_k =\frac{1}{e^{\beta_e(\ep_k -\ep_F)} +1}, ~~~\beta_e =\frac{1}{k_B T_e},
\eeq
where $T_e$ is the temperature of the electron sub-system (greater than the lattice temperature $T$ during the process of relaxation). $\ep_F$ is the Fermi energy. Free electron model $\ep_k = \frac{\hbar^2k^2}{2 m}$ was assumed where $m$ is the mass of an electron and $k$ is the magnitude of the electron wave-vector. For phonons, equilibrium Bose distribution was assumed:

\beq
n_q =\frac{1}{e^{\beta\hbar\om_q} - 1}, ~~~\beta =\frac{1}{k_B T},
\eeq
where $q$ is the magnitude of the wave-vector of an acoustic phonon mode (Debye model was used for phonon sub-system) and $\om_q = c_s q$ where $c_s$ is the sound speed for acoustic phonons. During the process of relaxation $T_e$ remains greater than $T$. Heat transfers from electron subsystem to phonon subsystem and then by the process of diffusion it goes out to the substrate or environment. 

Authors compute the amount of average energy transferred by electrons to lattice per unit volume and per unit time:

\beq
\bar{U} = \int d^3r \frac{d^3q}{(2\pi)^3} \dot{N}_q \hbar\om_q.
\label{3}
\eeq
Here $\dot{N}_q$ is the rate at which phonons are generated with wave-vector $q$ per unit volume. Each phonon carries energy of amount $\hbar\om_q$. Thus $\dot{N}_q \hbar \om_q$ is the amount of energy transferred (per sec per unit volume) to phonon modes with wave vector lying in the range $q$ to $q+dq$.  For the computation of $\dot{N}_q$ they use the Bloch-Boltzmann equation:

\begin{eqnarray}
\dot{N}_q&=&2 \int \frac{d^3k^\p}{(2\pi)^3} W_{k,k'} f_{k'} (1-f_k) [(n_q+1)\nonumber\\
&\times&\delta(\ep_{k'} -\ep_k -\hbar\om_q)- n_q \delta(\ep_{k'} -\ep_k +\hbar\om_q)].
\end{eqnarray}

\beq
W_{k,k'} = \frac{\pi U^2}{\rho V c_s^2}\om_q,~~~~\mathq = \mathk - \mathkp.
\eeq
 
Here $U$ is electron-phonon coupling constant, $\rho$ is the density of metal, $V$ is unit cell volume, $c_s$ is the sound speed, and $\om_q$ is the phonon frequency with wave-vector $q$. By using the expressions (1) and (2) for Fermi and Bose functions at different temperatures, and by imposing the assumption $\ep_F>>k_B T_e >> \hbar \om_q $ (generally valid for metals), the above equation  (equation 4) can be written as

\beq
\dot{N}_q=\frac{m^2U^2 \hbar\om_q}{2\pi \hbar^4 \rho V c_s}\frac{e^{\beta \hbar \om_q} -  e^{\beta_e \hbar \om_q}}{ (e^{\beta \hbar \om_q} - 1) (e^{\beta_e \hbar \om_q} - 1)}
\eeq

Technical details are given in\cite{nav1,pablo}. Using the Debye model for phonons, the average energy transferred by electrons to lattice per unit volume and per unit time (equation (\ref{3})) can be written as

\beq
\bar{U} = A\left\{ \left(\frac{T_e}{T_D}\right)^5 \int_0^{\frac{T_D}{T_e}} \frac{x^4 dx}{e^x -1} - \left(\frac{T}{T_D}\right)^5 \int_0^{\frac{T_D}{T}} \frac{x^4 dx}{e^x -1} \right\},
\eeq
where $A = \frac{2 m^2 U^2 (k_B T_D)^5}{(2\pi)^3 \hbar^7\rho c_s^4}$. The above expression simplifies in the following special cases:

CASE A: In the low temperature limit $T, ~T_e<< T_D$ the above expression gives:
\beq
\bar{U} = B (T_e^5 -T^5), ~~~~~B = A \int_0^\infty \frac{x^4 dx}{e^x -1}.
\eeq
In 2D, instead of $T^5$ it is $T^4$ behaviour\cite{nav101}. In a further sub-case $(T_e-T<< T<<T_D)$, we get
\beq
\bar{U} = \frac{2\pi^2}{3}\frac{m c_s^2 n}{\tau(T)}\frac{T_e -T}{T}.
\eeq

CASE B: In the high temperature limit $T,~T_e>>T_D$, equation (7) leads to 
\beq
\bar{U} = \alpha (T_e-T),~~~~~\alpha = \frac{A}{4T_D}.
\eeq
And in a further sub-case $(T_e-T<< T, ~~ T>>T_D)$, we get
\beq
\bar{U} = \frac{\pi^2}{6}\frac{m c_s^2 n}{\tau(T)}\frac{T_e -T}{T}.
\eeq
Here $\frac{1}{\tau(T)}$ is the equilibrium relaxation rate due to electron-phonon scattering as it appears in the theory of resistivity of metals (in the Bloch-Grueneisen formula)\cite{nav,ziman}.

\section{Two-Temperature Model (TTM) and laser excitation}

The idea of the study of TTM using laser excitation originated in 1973. S. I. Anisimov, B. L. Kapeliovich, and T. L. Perel'man\cite{ani} pointed out that when a metal surface is exposed to a picosecond laser pulse, emission current pulse (due to ejected electrons) from the surface of the metal is practically un-delayed relative to the laser pulse. This is due to small specific heat of electrons leading to preferential heating of it and during the course of the laser pulse (over picosecond time scales) electrons remain practically thermally insulated from lattice. This preferential heating of electrons leads to thermionic emission current pulse. That is, the thermionic emission of electrons is possible because heat absorbed by electrons from the laser pulse remains in the electron sub-system for short time scale of the order of psecs. They underlined that by measuring the thermionic emission over an extended timescale electron-lattice relaxation kinetics can be investigated. Laser pulses at that time were not short enough (not in the femto-second regime) and the study of electron-phonon relaxation kinetics remained an open area of research for some time.  It is interesting to re-visit their argument regarding separation of timescales.

The argument of the authors\cite{ani}, that the energy absorbed from the laser pulse mostly remains trapped inside the electron sub-system over a picosecond timescale,  goes as follows. They refer to equation no. 9 in KLT paper\cite{klt} (equation no (10) in the above section) and estimate heat transfer coefficient $\alpha$ between electrons and lattice:

\beq
\bar{U} = \alpha (T_e-T), ~~~\alpha = \frac{m^2U^2(k_BT_D)^5}{2 (2\pi)^3\hbar^7\rho c_s^4 T_D}.
\eeq

They estimated the value of $\alpha\sim 10^{17}erg/cm^3/sec/deg$. A typical heating time for phonons can be estimated as $\sim \frac{C_i}{\alpha}$ where $C_i$ is the phonon heat capacity. It turns out that this time scale is of the order of 100 psecs (this is an order of magnitude greater than the heating time for electrons and it translates to the fact that electronic heat capacity is about two orders of magnitude smaller than that of the lattice). Those ``hot" electrons which are not ejected out due to thermionic emission will transfer their energy to lattice via electron-phonon scattering.  Thus the authors argued that the evolution of disequilibrium of electrons (or anomalous heating of electrons as called in the literature in the 1980s) can be studied. But, It turns out that pico second laser pulses are not sufficient to observe the anomalous heating of the electrons. An experiment in 1983 made it clear (next section).

However, this sets a foundation for future experiments with shorter pulses (fsecs) to study preferential heating of electron sub-system and subsequent electron-phonon relaxations kinetics.

\section{First experiments that showed that electrons can be selectively excited using ultrashort laser pulses}

The first attempt to observe preferential heating of electrons (also called anomalous heating) was made by G. L. Eesley in 1983\cite{ees}. He used what is called Transient Thermomodulation Spectroscopy (TTMS)  which is an early version of  the pump-probe spectroscopy. He used 645 nm (1.92 eV) heating pulse (pump pulse) from a dye laser with temporal width of 8 psec and the sample used was 400 nm copper film. The pump pulse heat up the electrons and this further change the reflectivity of the copper film. The changed reflectivity was measured by a time-delayed probe pulse of similar width (8 psec) but weaker intensity. The probe pulse was time delayed with respect to the pump pulse and transient relative reflectivity change ($\Delta R/R$) was measured as a function of the time delay between the pump pulse (heating pulse) and the probe pulse. G. L. Eesley argued that anomalous heating of the electron sub-system is observed\cite{ees}.

But, time resolution was very low (pulse width was 8 psec). The data was mostly effected by the equilibrium heating of the electrons (it turns out that electrons and lattice remained in equilibrium on the time scale probed). And the temperature difference between electrons and lattice was less than a few Kelvin! However, G. L. Eesley noticed the need for femto-second laser pulses to differentiate anomalous heating of the electron sub-system and to study the kinetics of the electron-phonon relaxation. We quote his words here\cite{ees}:

\begin{quote}
 ``Extension of the technique into the femto-second regime should provide the capability to measure directly hot-electron relaxation times as a function of probe photon energy and as a function of both the transient and the equilibrium sample temperatures."
\end{quote} 

The first experimental observation of the anomalous heating (non-equilibrium electron distribution) came in 1984\cite{fuji}. J. G. Fujimoto, J. M. Liu, E. P. Ippen, and N. Bloembergen using 75-femto-second optical pulses demonstrated that electrons can be selectively excited (anomalous heating) and the electron-phonon relaxation happens on a time-scale of $~1$ psecs. The authors observe thermally enhanced photoemission from a tungsten metal surface. The key to the observation of anomalous heating is that the transient electron heating due to the pump pulse enhances the photoemission signal induced by the second probe pulse. By varying the time delay between pump and probe pulses,  time evolution of the selectively heated electron distribution can be studied. The estimated time scale for non-equilibrium electron distribution is found to be several hundred fsecs and the electron-phonon coupling constant is estimated to be of the order of $10^{17}~ergs/cm^3/sec/K$ which agrees with the estimate of S. I. Anisimov, B. L. Kapeliovich, and T. L. Perel'man\cite{ani}.

The experiment of 1984\cite{fuji} opened the floodgates for the studies of the anomalous heating of electrons in metals. Several results appeared in the late 1980s and early 1990s\cite{ultra,elsa,scho,bror}.

As the field advanced, the questions asked also became sharper: On what time scale the non-equilibrium electron distribution (non-Fermi-Dirac distribution) goes to the FD distribution via electron-electron scattering? How does the electron-phonon scattering effect the relaxation within the electron sub-system?  Whether phonons always remain in equilibrium during the process of electron-phonon relaxation? etc.

In 1987, this field saw an extension in a very novel way (next section).

\section{Using electron-phonon relaxation to investigate $\lambda$ (an important parameter in superconductivity)}

In 1987, P. B. Allen revisits the TTM problem posed by M. I. Kaganov, I. M. Lifshitz, and L. V. Tanatarov. In a seminal work\cite{allen} he generalizes the TTM in two important ways: (1) Instead of quadratic dispersion ($\ep_k \propto k^2$) valid for simple metals (as used by KLT), Allen generalizes the KLT calculation for an arbitrary dispersion $\ep(k)$; and (2) Allen  expresses $\alpha$ (refer to equation (12)) in terms of a very important parameter used in superconductivity theory ($\lambda\la \om^2\ra$):

\beq
\frac{dT_e}{dt} = \alpha (T-T_e), ~~~~~\alpha = \frac{3\hbar\lambda \la\om^2\ra}{\pi k_B T_e}.
\eeq

We briefly review this pioneering contribution. Allen uses the same set of physical assumptions as used by KLT:

\begin{enumerate}
\item The electron-electron (Coulombic) scattering is effective in maintaining a local equilibrium distribution of electrons (Fermi-Dirac distribution at an elevated temperature ($f_k$)) and phonon-phonon (anharmonic) scattering is also assumed to be effective in maintaining a local Bose-Einstein distribution for phonons ($n_q$).
\item Energy relaxation from electron sub-system to phonon sub-system is due to electron-phonon scattering (no other scattering is present there).
\item Diffusion due to spatial inhomogeneities is not present.
\item pump pulse is assumed to be a delta function in time (no light-matter interaction after $t=0^+$).  
 \end{enumerate}
 
 The evolution of the distribution functions is given by the Bloch-Boltzmann-Peierls kinetic equations:

\begin{widetext}
\begin{eqnarray}
\frac{\pr f_k}{\pr t} &=& \frac{2\pi}{\hbar}\frac{1}{N_c}\sum_q |M_{kk^\p}|^2  ( f_{k^\p}(1-f_k) [(n_q+1)\delta(\ep_{k'}-\ep_k -\hbar\om_q) + n_q \delta(\ep_{k'}-\ep_k +\hbar\om_q) ] \nonumber\\
&-&    f_k(1-f_{k'}) [(n_q+1)\delta(\ep_{k}-\ep_{k'} -\hbar\om_q) + n_q \delta(\ep_{k}-\ep_{k'} +\hbar\om_q)  ] )  \\
\frac{\pr n_q}{\pr t} &=& \frac{2\pi}{\hbar}\frac{1}{N_c}\sum_{k'} |M_{kk^\p}|^2   f_{k^\p}(1-f_k) [(n_q+1)\delta(\ep_{k'}-\ep_k -\hbar\om_q) - n_q \delta(\ep_{k'}-\ep_k +\hbar\om_q) ] \nonumber\\
&+& \frac{2\pi}{\hbar}\frac{1}{N_c}\sum_{k} |M_{kk^\p}|^2  f_k(1-f_{k'}) [(n_q+1)\delta(\ep_{k}-\ep_{k'} -\hbar\om_q) - n_q \delta(\ep_{k}-\ep_{k'} +\hbar\om_q)  ] .
\end{eqnarray}
\end{widetext}

The first equation in the above  array gives the time evolution of the thermal occupancy of electrons in the $k$th state. The scattering of electrons from $k'$ state to $k$ state and vice-versa along with the emission and absorption of phonons is written out and the corresponding matrix element of scattering is given by $M_{kk'}$. $N_c$ is the number of unit cells in the sample. The energy content of the electron sub-system and the phonon sub-system is given as

\begin{eqnarray}
E_e(t) &=& 2 \sum_k \ep_k f_k(t) \simeq E_0 +\frac{1}{2}\gamma_e T_e^2(t).\\
E_l(t) &=& \sum_q \hbar \om_q n_q  \simeq 3 N_a k_B T_l(t).
\end{eqnarray}

It can be easily verified that the total energy is conserved $\frac{d}{dt} (E_e(t) +E_l(t)) = 0$. Allen introduces the electron-phonon spectral function:

\beq
\alpha^2F(\ep,\ep',\Omega) \propto \sum_{k,k'} |M_{k,k'}|^2 \delta(\om_q-\Omega)\delta(\ep_k-\ep)\delta(\ep_{k'}-\ep').
\eeq
Where $\mathq = \pm (\mathk-\mathkp)$. By differentiating equation (16) w.r.t. time and after some relabeling of dummy variables in equations (14) and (15) with some algebra\cite{allen}, one obtains

\beq
\frac{dE_e(t)}{dt} = 2\pi N_c N(\ep_F)\int_0^\infty d\Omega\alpha^2 F(\Omega) (\hbar \Omega)^2 [n(\Omega, T_l) - n(\Omega, T_e)].
\label{main1}
\eeq
Here $n(\Omega, T_l) $ and $ n(\Omega, T_e)$ are Bose functions at lattice temperature $T_l$ and electron temperature $T_e$. Compare the above equation with equations (6) and (7) of KLT. No quadratic form of the electronic dispersion is used. Further, Allen introduces the moments of the electron-phonon spectral function:

\beq
\lambda \la \om^n \ra = 2 \int_0^\infty d\Omega \Omega^n \frac{\alpha^2 F(\Omega)}{\Omega}
\eeq

And in the high temperature limit $\frac{\hbar\Omega}{T_{l,e}}<<1$, the main equation (equation (\ref{main1})), on keeping the leading order terms,  leads to 

\beq
\frac{d T_e(t)}{dt} = \alpha (T_l-T_e).
\label{allen10}
\eeq 
Here 

\beq
\alpha = \frac{3\hbar\lambda\la \om^2\ra}{\pi k_B T_e}.
\eeq

This is a very important result (Compare equations (\ref{allen10}) and  (22) with  (12)). The conclusion is that by doing pump-probe spectroscopy a very crucial parameter needed in the theory of superconductivity $\lambda\la \om^2\ra$ can be estimated!

It turns out that in 1990 (about three years after Allen's work\cite{allen}) in a very crucial experimental work\cite{bror2}, S. D. Brorson and collaborators verified the predictions of Allen. The authors\cite{bror2} performed systematic pump-probe measurements on thin films of Cu, Au, Cr, Ti, W etc and estimated the electron-phonon coupling constant $\lambda$ using Allen's equation (\ref{allen10}). The agreement with other measurements of $\lambda$ was excellent.

These investigations showed that the basic assumptions in TTM are valid and some sort of quasi-equilibrium exists in electronic sub-system and phononic sub-system after photo excitation. But further investigations unfolded a different story and an apparent paradox arose in the field.

\section{Experiments that showed that TTM fails}

\begin{figure}[h!]
    \centering
    \includegraphics[width=\columnwidth]{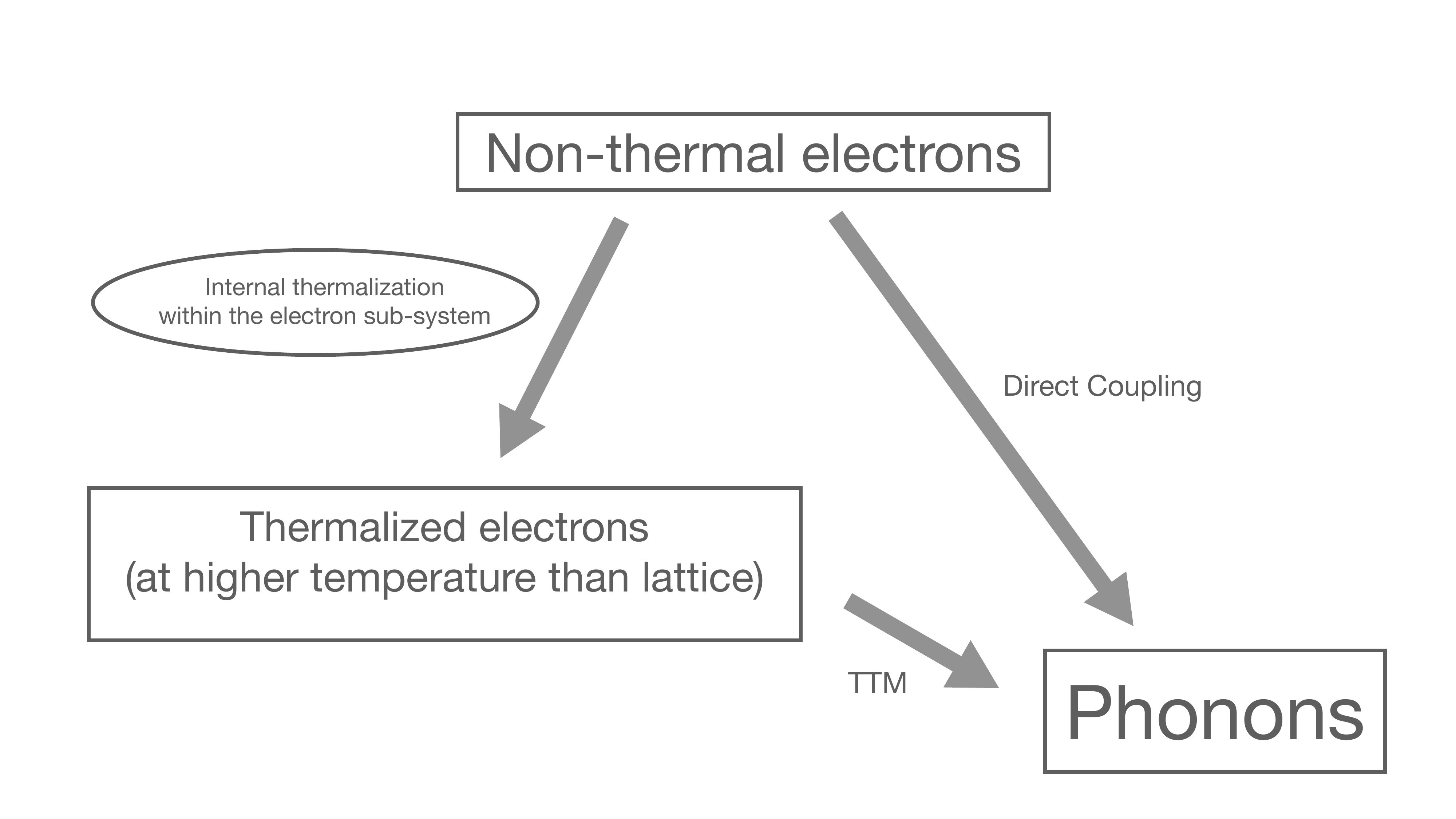}
    \caption{Three sub-systems: (1) nonthermal (non-FD distributed electrons); (2) Thermalized electrons (FD distributed electrons) at a higher temperature, and (3) Phonon subsystem.}
    \label{three}
\end{figure}

The basic assumption of the TTM (thermalized Fermi-Dirac distribution at an elevated temperature) makes sense when the internal relaxation time of the non-thermal electron distribution ($\tau_{e-e}$) is much less as compared to electron-phonon relaxation time $\tau_{e-ph}$ (that is, $\tau_{e-e}<<\tau_{e-ph}$). Investigations in the early 1990s\cite{fann1,fann2,sun1,rogier} showed that this is not true in general. In fact, it was estimated that  a non-thermal electron distribution takes about 500 fsecs to relax to a hot Fermi-Dirac distribution whereas $\tau_{e-ph}\sim 1~psec$ (in the case of polycrystalline gold films). Thus instead of two sub-systems one must consider three sub-systems (figure (\ref{three})). 

In 1992, W. S. Fann and collaborators\cite{fann1} used ultrafast photo-emission spectroscopy (instead of transient reflectance spectroscopy). They used 400 fsec visible (1.84 eV) heating pulse to create a non-thermal electron distribution in a gold film. For photoemission of this non-thermal distribution they used 700 fsec probe pulse. Although time resolution was low, they observed that, for time delays between pump (heating) pulse  and probe pulse of 400 fsecs, electron distribution substantially differs from a hot Fermi-Dirac distribution (they carefully take into account density of states factors\cite{fann1}). These observations clearly pointed to non-thermal electron distribution.

In another investigation by C. -K. Sun and collaborators\cite{sun1} different technique was used. They used transient reflectivity and transmissivity measurements in a pump-probe set-up. The authors used 140 fsec pump pulse in low fluence limit such that the electron temperature rise was only about 20 K. Also the pump pulse central wavelength was in the infrared regime so that only the intra-band excitation of electrons around the Fermi surface was probed. The probe pulse was 210 fsec and in it was in the visible regime. The observed transient reflectivity and transmissivity showed fast rise time and slow decay time behaviour (figure (3)).
\begin{figure}[h!]
    \centering
    \includegraphics[width=\columnwidth]{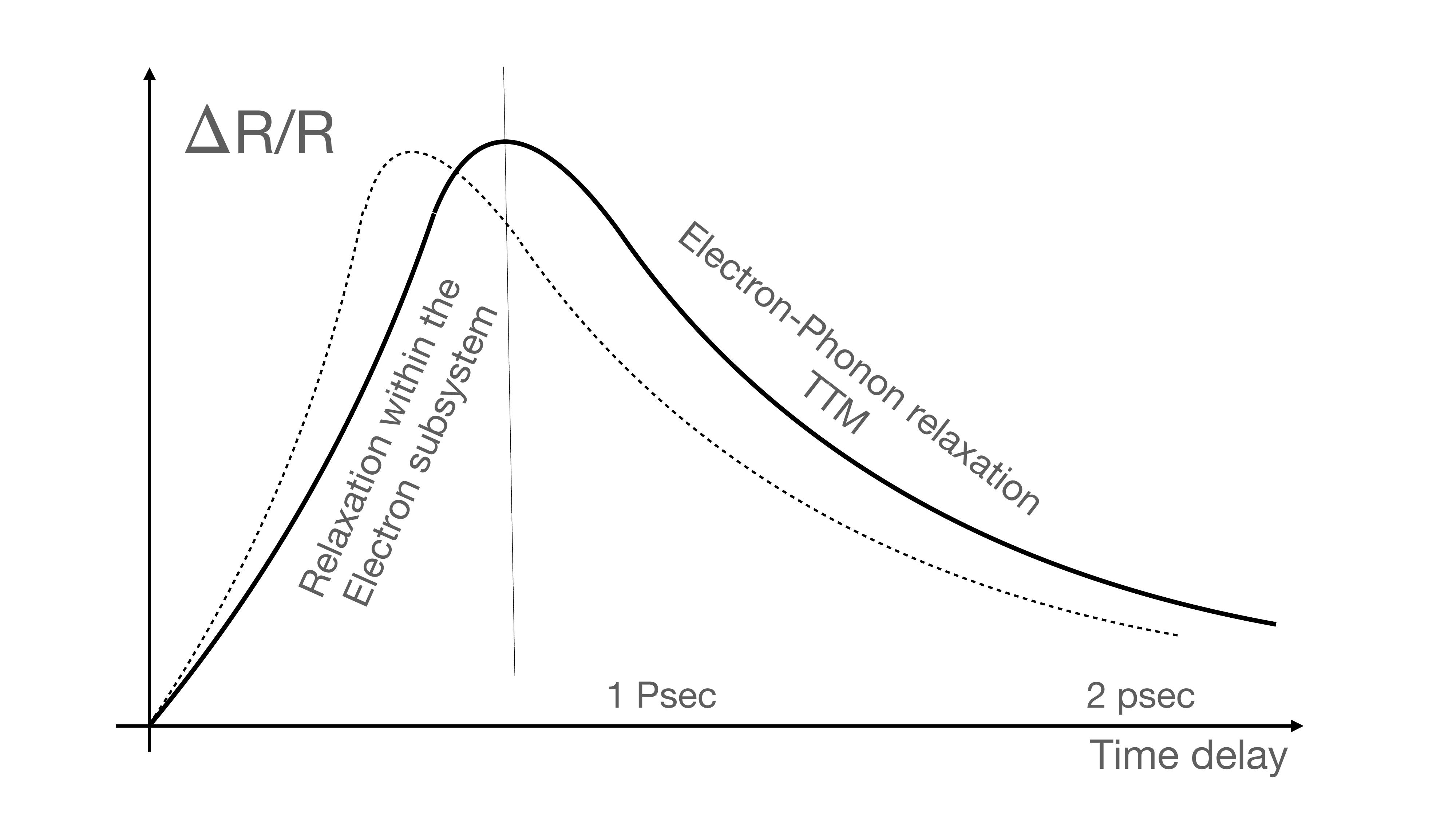}
    \caption{Signatures of non-thermal electron distribution (Schematic diagram. For original, refer to\cite{sun1}).}
\end{figure}

The authors first tried to reproduce the experimental data with the following single time-scale ($\tau_{el-ph}$ only) response function:
\beq
S(t) = \Theta(t) (e^{-t/\tau_{e-ph} +\alpha}),
\eeq
in which instantaneous thermalization of the electron gas is assumed and only electron-phonon relaxation is incorporated using single relaxation time $\tau_{el-ph}$. To account for the finite duration of the pulses, the response function was convoluted with measured pump-probe correlation function\cite{sun1}. The resultant graph is given by the dotted line in figure (3). The agreement is very poor.  Next, the authors included in the response function the rise time for a thermalized (hot) Fermi-Dirac distribution (that is, relaxation of a non-thermal electron distribution to thermalized (hot) Fermi-Dirac distribution):
\beq
S(t) = \Theta(t) (1-e^{-t/\tau_{e-e}}) (e^{-t/\tau_{e-ph} +\alpha}).
\eeq
This updated response function reproduced the data very well (both lines overlap very well, solid line in figure (3)). From this, the authors estimated that non-thermal electrons takes about 500 fsec to relax to a thermalized electron distribution (hot Fermi-Dirac distribution).

These observations showed that the simple minded two-temperature model is not sufficient to address the real state of affairs in a photo excited metallic sample.

\section{Beyond TTM}

In a pioneering experimental investigation in 1995 by Rogier H. M. Groeneveld and collaborators\cite{rogier} it is shown that electron distribution after photo-excitation remains a non-thermal (non-Fermi-Dirac) distribution on the time scale of electron-phonon relaxation. Thus the assumption that non-equilibrium electrons reach to an equilibrium (hot Fermi-Dirac) distribution on a time scale much smaller than electron-phonon relaxation is found to be invalid (at least in the low fluence limit\cite{rogier}). The authors used the expression

\beq
\Delta R(t) = (at+b)\Delta U_i(t) +c \Delta U_e(t),
\eeq

which is based on the theory by Rosei\cite{rosei}. Here $\Delta R(t)$ is the change in reflectance. Time dependence of internal energies ($\Delta U_i, ~~\Delta U_e$) is given by the approximated coupled equations:

\beq
\frac{dU_e(t)}{dt} = \frac{d(\frac{1}{2}\gamma T_e^2)}{dt} = \gamma T_e \frac{dT_e}{dt} = -\alpha_{T_i} (T_e(t) -T_i(t))
\eeq

\beq
\frac{dU_i(t)}{dt} = \frac{d(C_i T_i)}{dt} = C_i \frac{dT_i}{dt} = \alpha_{T_i} (T_e(t) - T_i(t))
\eeq

By considering $T_e(0), a, b ,c,~\alpha_{T_i}$ as fitting parameters, a very good fit is obtained with experimentally determined reflectance up to 10 psec. Initial temperature of the ions $T_i(0)$ is an experimentally known quantity from the thermometer attached to the sample. It was found that the fitting parameters $\alpha_{T_i}$ and $T_e(0)$ were largely determined by initial relaxation (from 0.25 psec to 4 psec) and are decoupled from lattice parameters ($a,b,~c$).

To make a tight comparison with the predictions of TTM, authors use its expression valid in the perturbative regime $T_e-T_i<<T_i$ (valid in the given experimental setup):

\beq
\frac{dU_e(t)}{dt} = \frac{d(\frac{1}{2}\gamma T_e^2)}{dt} = \gamma T_e \frac{dT_e}{dt} = -\alpha{T_i} (T_e(t) -T_i(t)),
\eeq

\beq
\frac{dU_i(t)}{dt} = \frac{d(C_i T_i)}{dt} = C_i \frac{dT_i}{dt} = \alpha{T_i} (T_e(t) - T_i(t)),
\eeq

where,

\beq
\alpha(T_i) = \frac{f(T_i)}{dT_i},~~~f(T_i) = 4 g_\infty (T/\Theta_D)^5\int_0^{\frac{\Theta_D}{T}} \frac{x^4}{e^x-1}dx
\eeq

TTM predicts that $\alpha(T_i) \simeq g_{\infty}$ when $T_i \agt\Theta_D$. By fixing the lattice temperature at 300 K, which is greater than Debye temperature for gold ($\Theta_D = 170K$), authors determine the coefficient $g_{\infty}$ by the same fitting procedure. Now one has all the information required to use TTM. However, when the lattice temperature was fixed at 100 K, authors found serious disagreement between the predictions of TTM and the experiment (refer to figure (4) in\cite{rogier}). The disagreement was seen at various fluence levels (still in the perturbative regime) and detailed discussions were presented on this aspect\cite{rogier}. Authors went further, they defined "instantaneous energy relaxation time":

\beq
\tau_E(T_e,T_i) = \frac{U_e(\infty)- U_e(0)}{dU_e/dt}
\eeq

This time scale was obtained both from the experiment and from the TTM, and serious disagreements were found. In conclusion, TTM is fund to be invalid in the low fluence limit (perturbative regime  $T_e-T_i<<T_i$). We notice that these investigations raise serious doubts on the validity of TTM.

Next, in an important theoretical investigation in 2002 \cite{reth} B. Rethfeld and collaborators pointed out a very curious aspect of non-equilibrium electron relaxation in metals. They considered energy absorption from laser field, electron-electron thermalization, electron-phonon thermalization all within the full Boltzmann collision integrals approach without using any relaxation time approximation. Detailed calculations are done for the case of Aluminum. The central result of this investigation can be expressed in the following way: For laser excitations near the damage threshold of the metal, the energy transfer from the non-equilibrium electrons to phonons can be expressed via the TTM, equation (10),  even when the perturbed electron distribution is very far from the hot Fermi-Dirac distribution!  Whereas in the regime of low laser excitation, relaxation is not according to TTM. It is much more delayed. It turns out that hot Fermi-Dirac distributed electrons are much more efficient in transferring energy to lattice than a non-equilibrium distributed electrons. These theoretical and simulation results corroborate the experimental findings of Rogier H. M. Groeneveld and collaborators\cite{rogier} in the perturbative regime.

{\it These investigations have a clear message: In the perturbative regime, actual relaxation is slow as compared to that predicted by the TTM. It implies that electrons do not reach to hot thermal equilibrium (Fermi-Dirac) distribution during the process of electron-phonon relaxation. If one assumes hot Fermi-Dirac distribution (like in TTM) we obtain mush faster relaxation in disagreement with the experiments. }


\section{An apparent Paradox and its resolution}

If TTM fails, how is it possible that the investigations of\cite{bror2} lead to a reasonably accurate estimate of the superconducting parameter $\lambda$ introduced into TTM by P. B. Allen? Allen's extension is based on TTM. 

It raises an apparent paradox! It turns out that in ref\cite{bror2} the experiments were done in the strong fluence limit  (not in the perturbative limit). In the strong fluence limit the relaxation proceeds roughly according to TTM as discussed in\cite{reth}. However, in the strong fluence limit, there is no proof that electron distribution reaches a thermal distribution (FD distribution) on a timescale much shorter than that of electron-phonon relaxation.  So, it seems quite surprising that measured value of $\lambda$ agrees very well with estimated value of it (from Allen's generalization of TTM).

\begin{figure}[h!]
    \centering
    \includegraphics[width=\columnwidth]{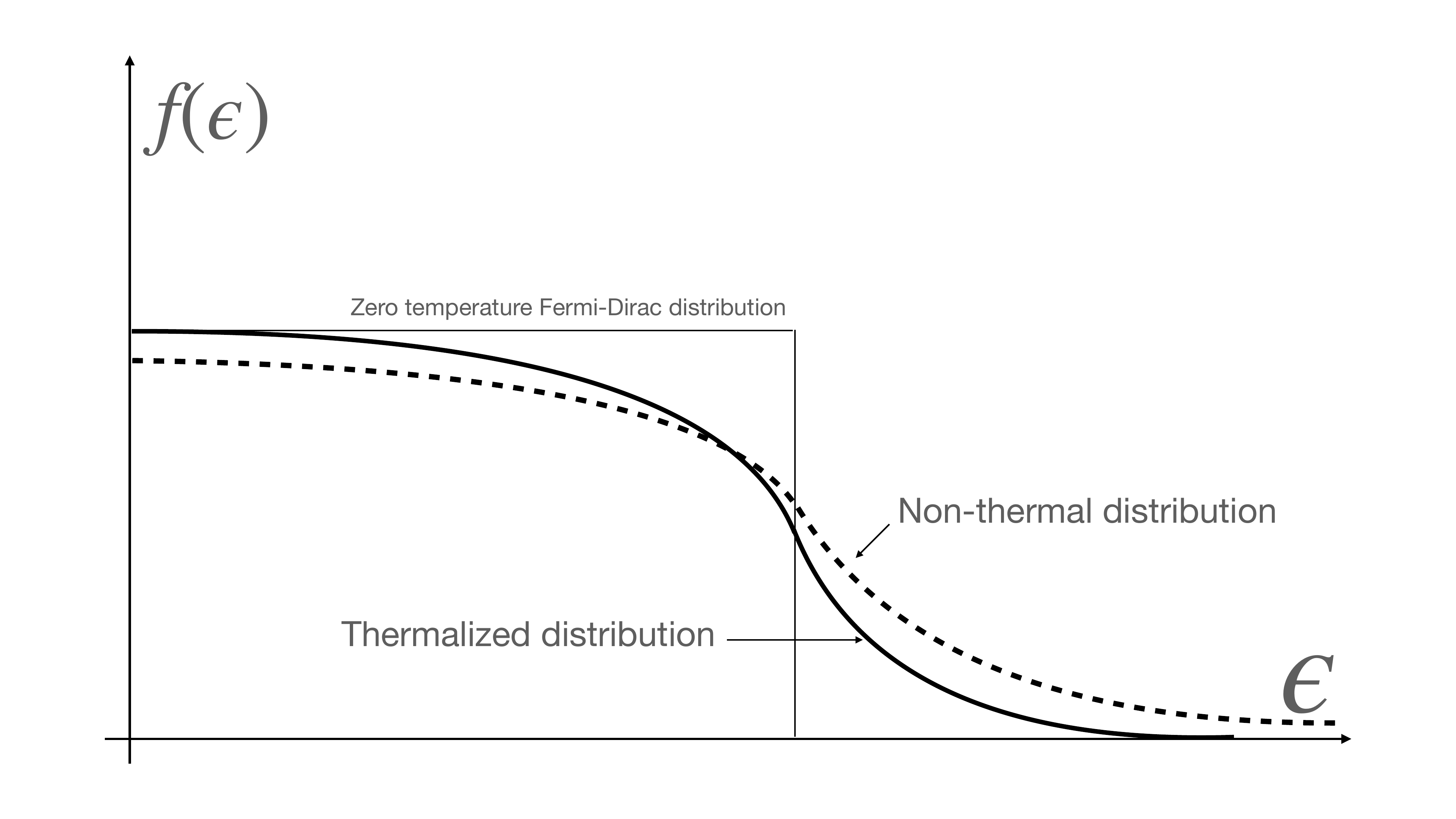}
    \caption{Two schematic electron distributions: one thermalized another non-thermal.}
    \label{twodis}
\end{figure}

It also must be noted that -- as far as the behaviour of some average property of the electron gas is concerned -- for a non-thermal electron distribution excited far from equilibrium and for a corresponding thermal distribution (that is with the same energy content) the energy relaxation may be considered to follow roughly some similar behaviour.  To be concrete, consider two schematic non-equilibrium electron distributions, one thermal and other non-thermal (figure (\ref{twodis})).  In a thermal distribution electron temperature can be defined, but $T_e>T$, whereas, in a non-thermal distribution electron temperature cannot be defined (however, energy content can be defined). There is more weight in the tail of the non-thermal distribution. But as far as some macroscopic parameter is concerned (such as $\lambda$ introduced by Allen in TTM) the relaxation behaviour of two distributions may be considered to be similar (without affecting the macroscopic results). However, as it stands, it is just a conjecture! Rigorous proofs, both experimental and theoretical are much needed.

\section{Summary of related investigations after year 2000}

In this section we briefly review some of the investigations which attempt to address the problems of TTM and advance ideas and theories that goes beyond it.

In year 2000, N. Del Fatti and collaborators\cite{del} also show that the assumption of almost instantaneous thermalization of non-equilibrium electrons is not valid. Authors measured internal thermalization timescale for non-equilibrium electrons in silver films and found it to be of the order of 350 fsecs. This is somewhat smaller than that for gold films ($\sim$ 500 fsecs). This difference is ascribed to reduced electron-electron screening in silver films as compared to that in gold films. These experiments were also done in the perturbative limit and corroborate the results of Lagendijk and collaborators\cite{rogier} (perturbative regime is defined as $T_e-T<<T$. Pump fluences of tens of $\mu J/cm^2$ typically leads to perturbative excitation of electrons in metals).

Next, it turns out that laser field also modifies the electron-phonon collision integral. In an important investigation\cite{andrey} Andrey V. Lugovskoy and Igor Bray takes into account this very effect (also konown as the Gurzhi mechanism\cite{gurzhi}). They also underlined the role of Umklapp electron-electron collisions. These two new inputs better accounted the experiments of Fann etal\cite{fann1, fann2}. The authors also concluded that field modified electron-phonon scattering rate is higher than that for electron-electron scattering. This means that electron-electron scattering is not sufficient to establish equilibrium within the electron subsystem during the process of the electron-phonon relaxation.

In 2006, E. Carpene\cite{car} extends the TTM by incorporating the initial non-thermal electron distribution within the relaxation time approximation. The author assumes the three temperature model (3TM): (1) a minority of non-thermal electrons, (2) majority of hot thermalized electrons, and (3) phonons. The non-thermal distribution ($\delta_{NT}$) is assumed to decay via electron-electron scattering and electron-phonon scattering considered within Relaxation Time Approximation (RTA):

\beq
\frac{\pr \delta_{NT}}{\pr \tau} = - \frac{\delta_{NT}}{\tau_{ee}} - \frac{\delta_{NT}}{\tau_{ep}}.
\eeq

The author computes the energy transferred from non-thermal distribution to the thermal distribution. Thus, for thermal distribution of electrons the non-thermal term (first term on the RHS of the above equation) acts like a heat source. Similarly, energy transferred to phonons from the non-thermal distribution (last term in the above equation) directly acts like a heat source for phonon distribution. Considering these physical features and by incorporating these into the original TTM, the author comes up with an updated TTM. Extensive numerical simulations exhibit marked deviations from TTM.

In 2017, Pablo Maldonado et.al.\cite{pablo} also considered three temperature model (3TM) just like that considered by Carpene\cite{car}. However, the phonon subsystem is analyzed in detail. Specifically, phonon modes with frequency $\nu$ and wave vector $q$ are taken to be interacting via  phonon-phonon interactions. Mode dependent "lattice temperature" is also defined. Phonon-Phonon interactions then lead to equilibrium in the phonon subsystem (that is the attainment of a single temperature for all modes). In addition, electronic and phononic heat capacities, electron-phonon and phonon-phonon linewidths were calculated ab initio (using DFT).  Extensive simulations for the system FePt showed that lattice takes about 20 psecs to reach equilibrium! The 3TM developed by the authors gives a reasonable material dependent description of relaxation phenomena in a given material without the need of any fitting parameters. 

Similarly, in the case of metal films, 3TM was introduced\cite{bezu}, where two sub-systems of phonons were considered. One set of phonons remain within the film, and in the other they cross the film-substrate boundary depending upon the angle of incident. Thus a concept of "leaky-phonons" is quite useful to study hot electron relaxation in metal films grown on substrates.

In 2017, TTM is extended to account for slow thermalization within the phonon sub-system in polar and non-polar semiconductors\cite{sri}.  In these systems electron-phonon and phonon-phonon interactions are very heterogeneous. It turns out that thermalization within the phonon sub-system (phonon-phonon interactions) acts like a "bottle-neck" (a limiting step)  for electron-phonon thermalization. Due to this effect, a single exponential decay of the electron-temperature (due to electron-phonon relaxation within TTM) changes to {\it multi-exponential decay} of electron temperature in polar and non-polar semiconductors (due to above mentioned heterogeneous interactions). This has very novel experimental consequences. Measurement of multi-exponential decay via pump-probe spectroscopy can provide a handle on the nature of heterogeneity of electron-phonon and phonon-phonon couplings and their spectral distributions. Refer also to\cite{yang} for non-equilibrium relaxation in semiconductors.

Very recently, Bethany R. de Roulet and collaborators\cite{rou}, using state of the art technology of attosecond science and the method of Attosecond Transient extreme ultraviolet light Absorption Spectroscopy (ATAS) showed that time scales of relaxation of nascent electron distribution after optical pump pulse can be measured to the finest accuracy available today.  Novelty of this technique lies in the fact that at such a small time scale (below 50 fsec) interference (perturbation) by phonons in the mechanism of relaxation of non-equilibrium electrons can be neglected.  It is mainly about the electron-electron interactions. In fact, the authors, using ATAS, show that non-equilibrium electron relaxation time scales in Mg, Pt, Fe, and Co are of the order of 38 fsec, 15 fsec, 4.2 fsec, and 2.0 fsec, respectively. It turns out that relaxation time scale matches remarkably well with the single electron life time given by the FLT:

\beq
\frac{1}{\tau} = A [(\pi k_B T_e)^2 + E^2] \simeq E^2.
\eeq

Here, the coefficient $A$ can be computed from the knowledge of EDOS and screened electron-electron scattering matrix element\cite{rou}. From the conditions of the experiment $k_B T_e<<E$, the last approximation in the above equation follows. 

It is quite counter-intuitive. Relaxation of a large number of non-equilibrium electrons is a many-body mechanism (should not be governed by single particle lifetime). But ATAS experiments and simple theory\cite{rou}  demonstrate that relaxation can be rationalized within the single -particle effects and the FLT. In author's opinion ATAS should be applied to a wider variety of materials where FLT is known to fail, such as strange metals. It will push the frontier in an entirely new direction. 
 
The crucial aspect that the author would like to underline is this: if the electron-electron relaxation time scale is below $50~ fsec$ in metals (as it is for the case of Mg, Pt, Fe, Co etc via ATAS) and the electron-phonon relaxation time scale is in the range of psecs, then, can one apply TTM to these systems? Answer is clearly yes! Then what about the investigations (with low resolution in the range of femto-seconds) of 1990s that showed TTM fails? In this author's opinion, all those old investigations should be re-visited with ATAS. Another crucial question would be: can one justify Allen's program with ATAS (that is, applicability of TTM)?

\section{Summary}

The famous Two-Temperature Model (TTM) used extensively in the investigations of energy relaxation in photo-excited systems originated in the seminal work of M. I. Kaganov, I. M. Lifshitz, and L. V. Tanatarov (KLT) in 1957. Then in 1974, S. I. Anisimov, B. L. Kapeliovich, and T. L. Perel'man pointed out that with an ultra short laser pulse a non-equilibrium state of electrons in metals can be created in which electron temperature is much greater than lattice temperature. This field experiences great developments in the 1980s and 1990s with the advent of femto-second (fsec) pump-probe spectroscopy.  The first experimental proof of this preferential heating of electrons ("anomalous heating" as it was then known) after photo-excitation was provided by J. G. Fujimoto, J. M. Liu, E. P. Ippen, and Bloembergen in 1984. In 1987, P. B. Allen re-visits the calculations of KLT and re-writes the electron-phonon heat transfer coefficient $\alpha$ in terms of a very important parameter in the theory of superconductivity ($\lambda \la \om^2\ra$). This has far-reaching consequences. Doing a pump-probe experiment, $\lambda$ for a given superconducting material can be estimated. However, as will be discussed in PART II of this review, the interpretation in the case of unconventional superconductors (such as cuprates) is non-trivial.  

In the early 1990s, it became very clear that the basic assumptions of the TTM fails (internal relaxation time of the non-thermal electron distribution $\tau_{e-e}$ is not short as compared to electron-phonon relaxation time $\tau_{e-ph}$ that is $\tau_{e-e}<<\tau_{e-ph}$). The first experimental proof of the non-equilibrium state of electrons (non-Fermi-Dirac distribution) was provided by several investigators including W, S. Fann and C.-K Sun and their collaborators. From year 2000 and onwards, focus has shifted from non-equilibrium phenomena in simple metals to that in strongly correlated systems such as high Tc cuprate superconductors and other unconventional superconductors. Very recently, with the advent of ATAS, we may be witnessing the coming back of TTM. But more investigations are needed. Some of the pressing issues are:  Why do in the low fluence limit ($T_e-T <<T$) experiments violate the predictions of the TTM? What are the role played by the long wavelength excitations in the electron gas (like plasmons). Other issues include: use of ATAS in the study of time evolution of the effect of exchange interactions (at atto-second and femto second time scales) for magnetic metals near their critical points, and ATAS should be used to check whether FLT is violated in strange metals at the initial stages of non-equilibrium electron relaxation.

\begin{acknowledgments}
Author dedicate this manuscript to the loving memory of Prof. N. Kumar (1 February 1940 -- 28 August 2017) who guided the author through this field. Author is also grateful to Prof. R. Srinivasan (of Mysuru) for helping the author with calculations, guidance, and many discussions related to several aspects of this field. He would also like to thank Yaroslaw Bazaliy and Igor Tanatarov for correspondence. 
\end{acknowledgments}




\begin{thebibliography}{99}
\bibitem{histo} This brief historical account is based on\cite{baz}. For more details reader may refer to it and references therein.
\end{thebibliography}
\end{document}